\documentstyle{frass}

\input epsf

\begin{document}

\begin{frontmatter}

\title{
Calibration of the Sensitivity of \\
Imaging Atmospheric Cherenkov Telescopes \\
using a Reference Light Source}

\author{A. Fra\ss},
\author{C. K\"ohler},
\author{G. Hermann},
\author{M. He\ss},
\author{W. Hofmann}

\address{Max-Planck-Institut f\"ur Kernphysik, P.O. Box 103980,
        D-69029 Heidelberg, Germany}

\begin{abstract} 
The sensitivity of an Imaging Atmospheric Cherenkov telescope
is calibrated by shining, from a distant pulsed monochromatic
light source, a defined photon flux onto the mirror. The light
pulse is captured and reconstructed by the telescope in an
identical fashion as real Cherenkov light. The intensity of
the calibration light pulse is monitored via a calibrated
sensor at the telescope; in order to account for the lower 
sensitivity of this sensor compared to the Cherenkov telescope,
an attenuator is inserted in the light source between the 
measurements with the calibrated sensor, and with the telescope.
The resulting telescope sensitivities have errors of 10\%,
and compare well with other estimates of the sensitivity.
\end{abstract}

\end{frontmatter}

Imaging Atmospheric Cherenkov Telescopes (IACTs) have evolved into
the most powerful tool for the study of galactic and extragalactic
$\gamma$-ray sources in the TeV energy range~\cite{overview}. In
IACTs, a (frequently tesselated) reflector with areas between a 
few m$^2$ and almost 100~m$^2$ is used to image the Cherenkov photons
emitted by an air shower onto a camera consisting of photomultiplier
(PM) pixels. The elliptical shower image traces the longitudinal
development of an air shower. The long axis of the image points to
the image of the source. The shape of the image allows to distinguish,
to a certain degree, compact $\gamma$-ray induced showers from the
more diffuse cosmic-ray showers \cite{cr_rejection}. The
power of IACTs can be improved significantly by operating multiple
IACTs in a stereo mode, observing the same shower with several
IACTs in coincidence. Stereo imaging allows the unambiguous spatial
reconstruction of the direction of individual air showers with a precision of 
$0.1^\circ$ and better \cite{stereo1,stereo2,hegra_stereo}, and
therefore provides the best angular resolution of all tools in $\gamma$-ray
astronomy. 

As the field matures, emphasis is shifting from the 
simple detection of TeV $\gamma$-ray sources to the precise determination
of source fluxes and their spectra. The spectra contain important clues
both concerning the acceleration mechanisms in galactic and extragalactic
particle sources, and concerning the propagation of $\gamma$-rays and
their interaction in particular with extragalactic radiation fields.
One of the key issues in the measurement of fluxes and spectra is
the energy calibration of an IACT, i.e, the determination of 
the relation between the
signal size (measured, e.g., in units of ADC counts) and the incident
photon flux, or ultimately, the energy of the air shower. Because of
the steeply falling spectra, calibration errors are amplified in the
calculation of integral fluxes above a certain energy threshold.
Lacking a suitable monochromatic ``test beam'', the energy calibration
of IACTs has to be derived indirectly. Presently, uncertainties
in the energy calibration of IACT frequently range in the 
20\% to 30\% region and the resulting systematic errors
are the dominant term in measurements of the $\gamma$-ray flux.
In this paper, we describe 
a technique to calibrate, in one single step, the response of an IACT
and its readout chain. The paper is structured as follows: the next
section (1) contains a quantitive discussion of the IACT calibration issue,
and gives examples of calibration techniques. In the following
sections (2 and 3), our technique is introduced, 
and the implementation is described.
A final section (4) is dedicated to the discussion of the results
and error sources.

\section{Response and energy calibration of IACTs}

For the calibration of the response of an IACT, two
coefficients are relevant. The relation between light intensity 
(in photons/m$^2$) and
the digitized telescope signal (in ADC channels)
provides an overall scale factor
in estimates of shower energies. The signal per single photoelectron
is needed in addition to obtain the actual number of photoelectrons
in a given image, which determines the size of fluctuations around
the mean response. Scope of this paper is the determination of
the first of these two coefficients; methods to determine the
second are discussed e.g. in \cite{single_pe,npe,npe1}.

Usually, the total intensity $M$ of a Cherenkov image, given in
units of ADC channels, is used as a measure of the energy of a
$\gamma$-ray shower. The radial dependence of the intensity of the
Cherenkov light is taken into account by either selecting events with
impact points within a certain limited distance range, or by explicit
estimation of the impact distance and corresponding correction factors.
Let $I(\nu)$ be the intensity
\footnote{In the context of photon emission from air shower, we
use the terms ``photon flux'' and ``photon intensity'' as synonyms.}
(photons of frequency $\nu$ per unit
area) the air shower would have generated at the location of the
telescope without intermediate absorption or scattering in the 
atmosphere, and $T_{atm}(\nu)$ the atmospheric transmission (which
of course depends on the height distribution of photon emission).
The magnitude $M$ of the image can then be written in terms of an
IACT response $K_{IACT}(\nu)$,
\begin{equation}
M =  \int~I(\nu)~T_{atm}(\nu)~K_{IACT}(\nu)~\mbox{d}\nu~~~.
\end{equation}
The response $K_{IACT}$ includes the effective mirror area
$A_{eff}$ (taking into account shadowing by the camera, the
camera masts, etc.), the mirror reflectivity $R_{M}(\nu)$,
the efficiency of light collection onto the PMs (e.g. with
funnels etc.) $\epsilon_{LC}(\nu)$, the quantum efficiency
and photoelectron collection efficiency of the PM $\epsilon_{PM}(\nu)$,
the PM charge amplification $G_{PM}$, and the electronics
gain and digitizer conversion factor $G_{el}$:
\begin{equation}
K_{IACT}(\nu) = A_{eff}~R_{M}(\nu)~\epsilon_{LC}(\nu)~
\epsilon_{PM}(\nu)~G_{PM}~G_{el}~~~.
\label{eq_eff}
\end{equation} 
We assume here that the camera has been flat-fielded, such
that a common calibration constant can be applied for all PMTs
of the camera; the terms entering Eq.~\ref{eq_eff} then 
represent suitable averages.

Various techniques to calibrate IACT response have been used
or proposed.
One can, e.g., combine data-sheet specifications, educated
guesses, and measurements for the individual factors entering
$K_{IACT}$. In particular, the crucial factor $G_{PM} G_{el}$
can be determined by either directly observing the 
single-photoelectron peak in the digitized spectrum (see, e.g.,
\cite{single_pe}), or by determining the mean number of 
photoelectrons generated by a test light pulse on the basis of the relative
width of distribution of digitized signals, which is proportional 
to $1/\sqrt{n_{pe}}$, modulo corrections for the width of the
single-photoelectron peak, intensity fluctations of the test
light pulse, and electronics and digitizer noise; see, e.g., 
\cite{npe,npe1}.
A problem with this technique is that data-sheet specifications
are not always reliable (mirror reflectivity will e.g. deteriorate
over time), and that measurements of the individual factors are 
non-trivial, so that the combined errors are quite significant
(see section 4).

Other techniques aim at measuring either $K_{IACT}$ or 
$T_{atm} K_{IACT}$ directly. All these techniques determine
a spectrum-averaged calibration constant:
\begin{itemize}
\item Cosmic rays can be used to calibrate the overall 
response of the IACT, including the leading effects of
atmospheric transmission (see, e.g., \cite{cr_cal}). 
The small corrections in transmission between the
(deeper) hadronic showers and $\gamma$-ray showers are
derived from Monte-Carlo studies.
Technically, one compares the MC-predicted cosmic-ray counting
rate and the measured counting rate, and adjusts a global
calibration factor $K$ such that the two agree; for an integral spectral
index $\alpha$, the detection rate varies approximately like $K^{-\alpha}$.
One drawback of this technique is that the predicted rates
depend on the cosmic-ray flux and in particular on the composition
\cite{cr_cal,cr_flux}, and uncertainties in these quantities propagate into
the  $\gamma$-ray flux. Another difficulty is that 
one has to rely on the proper modeling of hadronic showers,
which involves larger uncertainties than in the case of
electromagnetic showers.
\item Cherenkov rings generated by local muons can be triggered
and reconstructed by telescopes with sufficiently large mirror
areas; the light yield is then used to calibrate the response
\cite{cherenkov_rings}. One difficulty is that only the 
spectrum-weighted average response is measured, and that the
spectrum of Cherenkov light from nearby muons contains increased UV
components compared to showers high up in the atmosphere.
\item Starlight from stars selected to match the Cherenkov spectrum
and giving rise to a DC current in the PMTs can be used for calibration
\cite{starlight}. Problems with this technique include the fact
that the electronics signal path is usually quite different for
 the measurement of DC currents and for fast pulses,
and that hence the relevant gain factors are not measured directly.
\end{itemize}

The technique discussed in this paper aims at measuring
$K_{IACT}(\nu)$ directly, by generating from a distant, monochromatic, 
pulsed light source a known flux of photons at the telescope,
which illuminates uniformly the entire mirror and
which is captured and digitized like a genuine Cherenkov light 
front.

\section{Calibration setup with a distant pulsed light source}

The key issue in the calibration of an IACT with a distant pulsed light 
source is the determination of the light flux at the location
of the telescope. For light pulses of ns-duration, no other
detector can compete in sensitivity with Cherenkov telescopes,
and hence there is no immediately suitable reference detector.
One option would be to calibrate the light source with a detector
closer to the source, using the $1/r^2$ dependence of the light flux.
In this case, however, atmospheric attenuation enters again.
A more promising option is to make the light source strong enough
that it can be detected by a calibrated sensor at the
telescope, and then insert a calibrated attenuator for the measurements
with the telescope itself. Of course, care must be taken that the 
attenuator influences only the intensity, but no other properties 
of the beam such as, e.g., its angular spread.

This calibration procedure was tested using one of telescopes
(CT4) of the HEGRA IACT system~\cite{hegra_stereo}, which
is operated
on the Canary Island of La Palma, 
at the Observatorio del Roque de los Muchachos
of the Instituto Astrofisico de Canarias. These telescopes
use tesselated 8.5~m$^2$ mirrors with 5~m focal length, and
are equipped with 271-pixels camera with a field of view of
$4.3^\circ$. The cameras are read out by flash-ADC digitizers.

As a calibration light source, 
we used as pulsed 337-nm Nitrogen laser to excite 
a scintillator (NE 111), which re-emits at wavelengths above
 350~nm. The pulse length generated by the laser
was 0.5~ns. The roughly isotropic light from the scintillator
was filtered through interference filters of 10~nm bandwidth. To attenuate
the light output, attenuators 
(neutral density filters) were inserted into the primary laser
beam, prior to the scintillator. The attenuation was calibrated by
monitoring the light output from the scintillator by a reference
photodiode. Attenuating the input pulse into the scintillator
(rather than its output) should guarantee that the spatial 
distribution of the re-emitted beam is invariant.

As sensor at the telescope, a calibrated large-area
(about 1~cm$^2$), low-capacitance
(and hence low-noise) photodiode was employed. Coupled to a low-noise
charge-sensitive preamplifier and a shaping amplifier, light pulses of
less than 10000 photons can be detected, and their intensity measured to 
good precision by averaging over a larger number of pulses.
 
The proper distance between light source and telescope deserves
some attention. With a light source ``at infinity'', the calibration
light is imaged onto a small spot of a single pixel, resulting
in two undesirable features: a) one is very sensitive to local
variations in photo efficiency, and b) because of the limited
dynamic range of a single pixel, the calibration light pulse 
should contain no more that a few 1000 photons incident on the
mirror. The latter condition implies a flux of a few $10^{-2}$
photons/cm$^2$. Between the calibration measurement and the 
measurement with the telescope, the light source would have to
be attenuated by a factor a few $10^5$. Such large factors are non-trivial
to measure with the required precision of a few \%. Therefore,
the light source was moved closer, to about 65~m from the telescope.
In this position, the focal plane is behind the camera, but
the image is still well contained
within the camera. Since the intensity is spread over O(100)
pixels, one can use larger intensities, and the
attenuation needed between the two measurements drops to a more
manageable factor of $10^3$ to $10^4$.

\begin{figure}[htb]
\begin{center}
\mbox{
\epsfxsize12.0cm
\epsffile{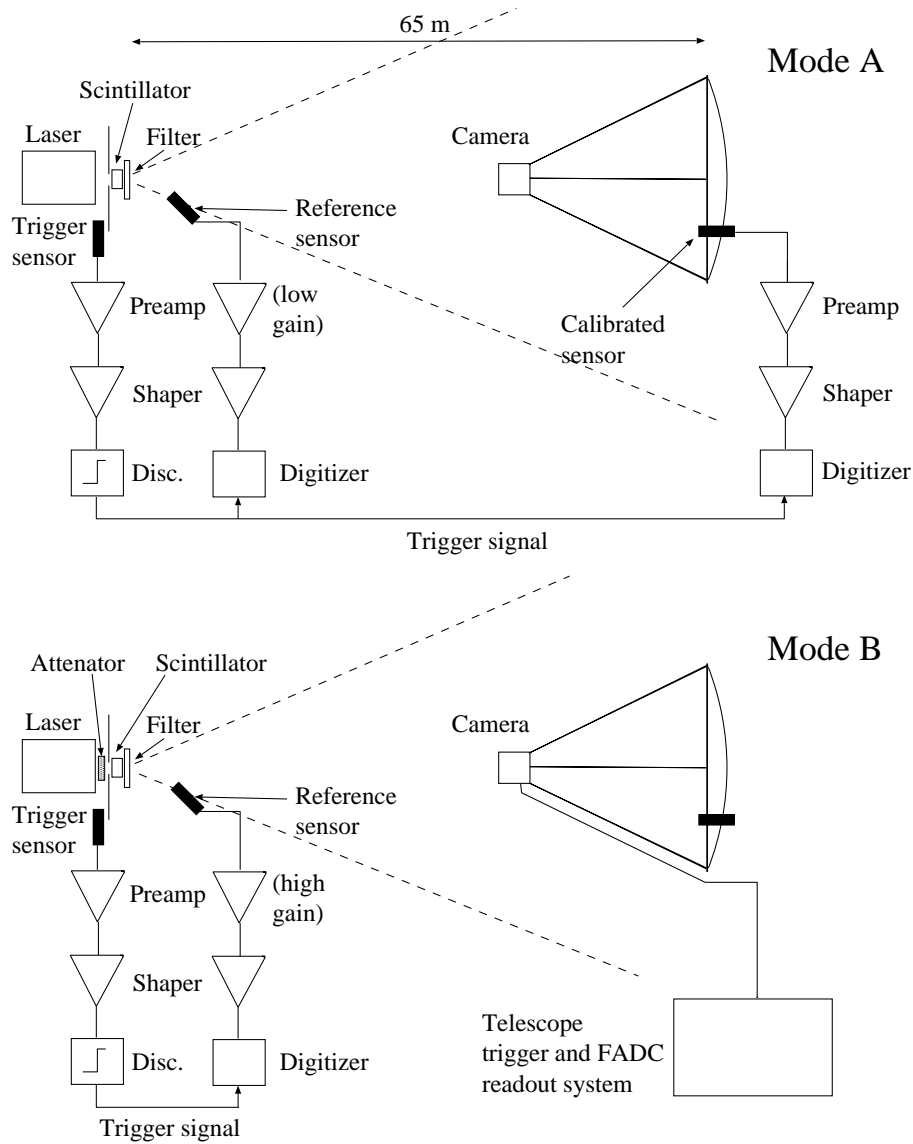}}
\end{center}
\caption
{Configurations used in mode A to calibrate the light
intensity at the telescope, in in mode B to measure the 
telescope response.}
\label{fig_setup}
\end{figure}

The actual setup is illustrated in Fig.~\ref{fig_setup}(a)
for the mode `A' where the light intensity at the telescope
was calibrated, and in Fig.~\ref{fig_setup}(b) for the mode `B'
where the telescope signal was measured. The laser was 
contained in a light-tight box, which separate compartments
for the attenuator (in mode B), and for the scintillator and 
interference filter.
The laser was pulsed at 4~Hz. A photo diode in the laser 
compartment was used to pick up some stray laser light and to
provide a trigger signal for the digitization of the other
diode signals. The light intensity was monitored by a 
reference photo diode (Hamamatsu S 3590-06) 
mounted at a distance of 31~cm from the
scintillator, close to (but not in) the beam to the telescope
mirror. In mode A, this diode was coupled to a low-gain 
charge-sensitive preamplifier, in mode B a high-gain preamplifier
was used, in each case together with a shaping amplifier with
3~$\mu$s shaping time. The signal of the calibrated photo diode
at the telescope (Hamamatsu S 3590-06) is similarly amplified. 
At a depletion voltage
of 28~V, the capacitance of this diode is 50~pF, resulting in a 
rms noise of the system of 550 electrons at room temperature,
and about 320 electrons under operating conditions, at a few
Deg. C, where in particular the leakage current of the diode
is strongly reduced. In mode B, where the
telescope is read out, the regular trigger system and data
aquisition system of the telescope is used to capture the light 
pulse. During the calibration measurements, the telescope will
in addition occasionally trigger on noise or cosmic rays. Such
events are easily removed during the offline analysis. Great care
was taken to properly align all elements involved in the
calibration, to eliminate stray light incident on the diodes
by baffles, and to minimise electronics noise and pickup
by proper grounding. Electronics pedestals and gains were
monitored regularly.

\section{Telecope calibration and systematic errors}

\begin{figure}[htb]
\begin{center}
\mbox{
\epsfxsize9.0cm
\epsffile{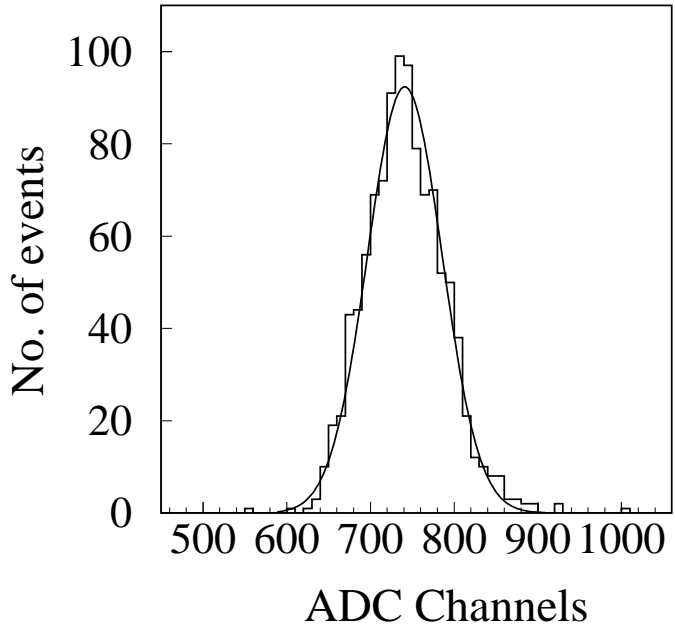}}
\end{center}
\caption
{Example for the distribution of pulse heights measured
with the calibrated sensor at the telescope in mode A.}
\label{fig_sensorsignal}
\end{figure}

\begin{figure}[htb]
\begin{center}
\mbox{
\epsfxsize11.0cm
\epsffile{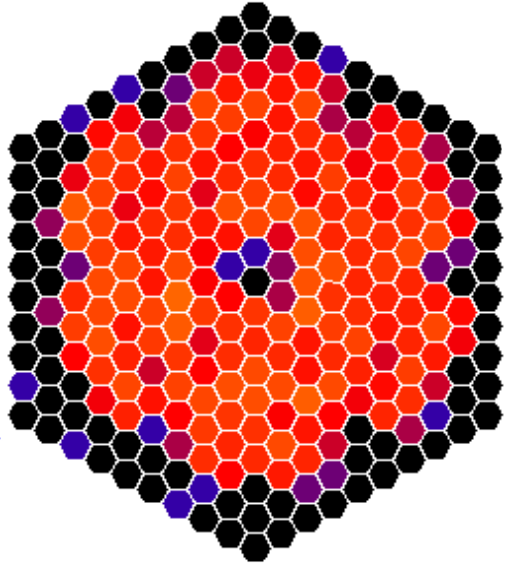}
}
\end{center}
\caption
{A typical calibration light pulse, as viewed by the telescope
camera in mode B. The illuminated pixels trace the somewhat
irregular outer contour of the tesselated mirror; the hole in
the center is caused by the shadow of the camera.}
\label{fig_camerapict}
\end{figure}

\begin{figure}[htb]
\begin{center}
\mbox{
\epsfxsize9.0cm
\epsffile{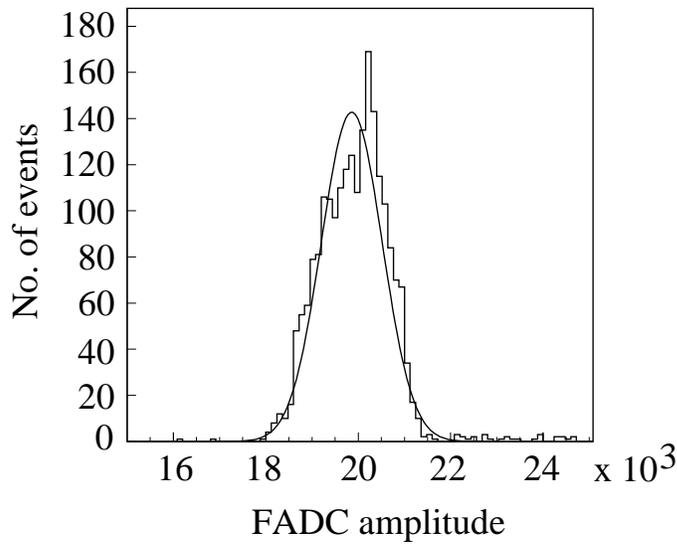}}
\end{center}
\caption
{Distribution of the total signals seen by the camera (in ADC units),
for a typical calibration run.}
\label{fig_cameradist}
\end{figure}

Calibration measurements were taken at two center wavelengths,
430~nm and 470~nm, during two measurement campaigns. Based on the
experience in the first campain, a number of sources of systematic
errors were identified and corrected. We therefore use only the
data from the second campaign. Poor weather conditions allowed only
three sets of calibration measurements, at 430~nm and at 470~nm 
in one night, and a second 430~nm measurement in another night.
In between, the setup was partly disassembled, so the comparison
of the two measurements at 430~nm serves to test the reproduceability
of the procedure. One measurement typically included 1000 or more
laser shots; statistical errors are therefore generally small, 
and the precision of the calibration is entirely dominated
by systematic effects. Figs.~\ref{fig_sensorsignal} through 
\ref{fig_cameradist} show characteristic
data from various steps of the procedure: the signal of the
calibrated sensor at the telescope in calibration mode A
(Fig.~\ref{fig_sensorsignal}), the typical image detected in the 
camera in mode B (Fig.~\ref{fig_camerapict}), and the intensity
detected in the camera (Fig.~\ref{fig_cameradist}) after
processing with the usual analysis chain. 

The telescope sensitivity $K_{IACT}(\nu)$ is derived in the
following steps:
\begin{itemize}
\item Using the photo diode calibration provided by the
manufacturer, and the calibrated sensitivity of the charge
amplification chain, the light flux in mode A is calculated
from the averaged signal.
\item Using the signals of the reference diode near the
light source in modes A and B, and the calibrated gains of 
the low-gain and high-gain amplification chains, the 
attenuation factor between mode A and B and hence the flux
at the telescope in mode B is calculated.
\item Dividing the average IACT signal in mode B (in units of
ADC counts) by the flux, the sensitivity $K_{IACT}$ is
derived.
\end{itemize} 
With this procedure, the parameters listed in Table~\ref{tab_results}
were derived.
The two measurements at 430~nm, taken in different
nights with a partial disassembly of the components in between,
yield results which agree within 1.4\%.
{\small

\begin{table} [htb]
\begin{center}
\begin{tabular}{|l|c|c|c|}
\hline
Wavelength [nm] & 430 (I) & 430 (II) & 470 \\
\hline 
Photons/pulse, Mode A [1/cm$^2$]
   & $9.90 \cdot 10^3$ & $9.48 \cdot 10^3$ & $2.99 \cdot 10^3$ \\
Attenuation                       
   &  $6.01 \cdot 10^3$ & $6.01 \cdot 10^3$ & $1.92 \cdot 10^3$ \\
Photons/pulse, Mode B [1/cm$^2$]
   &  1.65 & 1.58 & 1.56 \\
Total camera signal [ADC counts]
   & $1.99 \cdot 10^4$ & $1.88 \cdot 10^4$ & $1.39 \cdot 10^4$ \\
$K_{IACT}$ [ADC counts/(photon/cm$^2$)] 
   & $(1.21 \pm 0.12) \cdot 10^4$ & $(1.19 \pm 0.12) \cdot 10^4$ & 
     $(0.89 \pm 0.09) \cdot 10^4 $\\
$K_o$ [ADC counts/(photon/cm$^2$)]
   & $(1.13 \pm 0.25) \cdot 10^4$ & $(1.13 \pm 0.25) \cdot 10^4$ & 
     $(0.81 \pm 0.18) \cdot 10^4 $\\
\hline
\end{tabular}
\vspace{0.5cm}
\caption{Summary of the results of the calibration 
measurements, listing the mean number of photons per pulse
in modes A and B, the attenuation between A and B, the
average camera signals, the IACT sensitivity $K_{IACT}$
derived from these measurements, and the expected sensitivity
$K_o$ based on a priori knowledge of telescope properties,
and the measurement of the number of ADC channels per photoelectron
(see section 4).}
\label{tab_results}
\end{center}
\end{table}

}

Apart from the practically negligible statistical errors, the
following elements were considered as the major sources of systematic errors
in the determination of $K_{IACT}$:
\begin{itemize}
\item the sensitivity and effective area of the calibrated photo diode
at the telescope
\item differences in the conditions during diode calibration and 
during actual use
\item the absolute calibration of the amplification chain
for the telescope diode
\item the absolute calibration of the two (high-gain and low-gain)
amplication chains used for the reference diode to determine
the attenuation between modes A and B
\item deviations from linearity of the camera pixels for
large pulse heights
\item a slight asymmetry in the distribution
of IACT signals (Fig.~\ref{fig_cameradist})
\item the uniformity of the illumination of the mirror, and 
the slight changes in light paths due to the relative proximity
of the light source.
\end{itemize}

The calibrated photo diode (Hamamatsu S 3590-06) was 
specified with a sensitivity 
of 0.222~A/W at 430~nm, and 0.286~A/W at 470~nm, with errors
of $\pm 5\%$. The effective area of the diode was determined
by measuring the photocurrent for uniform illumination through
a series of diaphragms with different diameters, and entirely
without diaphragm. From the current with the fully illuminated
diode, and the measured curve of current vs. illuminated area,
the effective area $A$ of the diode was determined to $0.89 \pm 2$
mm$^2$. The relevant product of sensitivity times effective
area was checked against three other photo diodes (two Newport
818-UV/CM with $A = 1.00$~cm$^2$, calibrated to $\pm$ 2\% in the relevant
wavelength region, and a Gigahertz SSO-BL-50-2-BNC with 
$A = 0.50$~cm$^2$, calibrated to $\pm$ 4\%) by uniformly 
illuminating the diodes; the results were in good agreement 
within the quoted errors.

The photo sensor used to measure the photon flux at the
telescope is factory-calibrated at $25^\circ$C
under continuous illumination
without external bias.
For our measurements, it was operated with short (ns) light
pulses, and with a bias voltage of 28 V in order to deplete
the diode and to minimize detector capacitance and hence 
noise. Temperatures were around $3^\circ$C to $5^\circ$C.
Laboratory measurements showed that the application of the
bias voltage increased the diode output by 1\%. According
to specifications, the diode calibration should apply both
to DC mode (with a sensitivity expressed in A/W) and to
pulsed mode (sensitivity in C/J). We verified that the
sensitivity does not depend on the duty cyle of the light source,
within 2\% errors. Temperature dependence is more critical;
while no specifications were given for the diode used in the
setup, similar diodes show an increase of up to 4-5\% in output
at 430~nm, when the temperature is lower by $25^\circ$C compared
to the calibration, and a reduced effect at 470~nm. Laboratory
measurements showed indeed such a behaviour, within 2\% to 3\% errors.
Corresponding correction factors were applied.

\begin{figure}[htb]
\begin{center}
\mbox{
\epsfxsize9.0cm
\epsffile{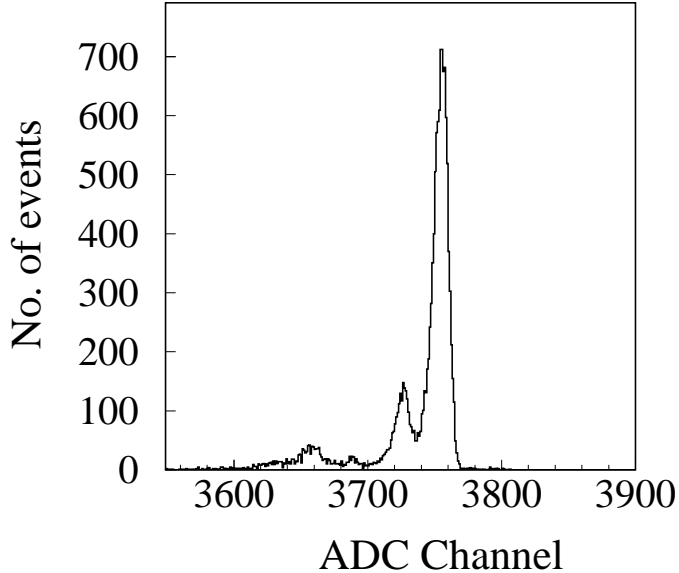}}
\end{center}
\caption
{Pulse height distribution deteced when irradiating the calibrated
photo diode with $\alpha$ particles from an Am source, with energies
of 5.443~MeV and 5.486 MeV.}
\label{fig_alpha}
\end{figure}

Two different techniques were used to calibrate the charge
amplifiers. One method was to inject a defined amount of charge
via a calibration capacitor. A problem with this technique is
that for small capacitance values the measurement of
the capacitance is non-trivial; for large capacitors errors may
be introduced since the input impedance of the amplifier can no
longer be neglected. We used capacitors between 2 and 20 pF,
measured to 5\% for $C < 10$~pF
 and to 3\% for larger values.
Calibration results with the different capacitors were consistent. 
A second calibration technique was to use the actual
photo detector to detected $\alpha$-radiation from $^{241}$Am decays,
thereby depositing a well-defined amount of charge in the
actual detection device, without requiring any additional external
elements. Source and detector were placed in a vacuum vessel to 
avoid energy loss in air. Fig.~\ref{fig_alpha} shows a typical
spectrum with two $^{241}$Am lines of 5.443~MeV and 5.486~MeV. 
The resulting amplifier gains
are listed in Table~\ref{tab_gains}, together with the values 
obtained with the charge injection via capacitor. For the two
high-gain amplifiers, both techniques give consistent results.
For the low-gain amplifier, the methods deviate by 6.5\%, or
1.5 standard deviations. No convincing explanation was found for
this discrepancy. For the calculation of telescope sensitivity,
we use the calibration constant obtained with the $^{241}$Am source,
but we enlarge the error to include the value obtained with the
capacitor calibration.
\begin{table} [htb]
\begin{center}
\begin{tabular}{|l|c|c|}
\hline
Gains in V/C & Capacitor charge injection & $^{241}$Am source \\
\hline 
High-gain amp., reference diode 
   & $(1.69 \pm 0.05) \cdot 10^{13}$ &  $(1.71 \pm 0.03) \cdot 10^{13}$ \\
Low-gain amp., reference diode
   & $(1.03 \pm 0.03) \cdot 10^{10}$ &  $(0.96 \pm 0.03) \cdot 10^{10}$ \\
High-gain amp., telescope diode
   & $(1.52 \pm 0.05) \cdot 10^{13}$ &  $(1.54 \pm 0.03) \cdot 10^{13}$ \\
\hline
\end{tabular}
\vspace{0.5cm}
\caption{Gains of the diode readout chains (preamplifier 
and shping amplifier), measured with two different calibration
techniques. The gains quoted refer to a low gain setting (`20')
of the shaper amplifier; in most of the actual measurements, a higher
(calibrated) gain (`1000') was used. }
\label{tab_gains}
\end{center}
\end{table}

The calibration procedure at 430~nm resulted in pulse
heights of some camera pixels of 180 and more photoelectrons. From
earlier studies of the linearity of the PM and the amplification
chain we know that at this pulse height deviations from linearity
cannot be neglected, caused mainly by nonlinearities in the PM
response. On average, pulse heights of those
high pixels are 3\% to 4\% lower than expected
for a linear response. A corresponding correction factor was 
applied, with an overall systematic error of 2\%.

We note further that the distribution of the integral pulse
height detected in the camera is not exactly symmetric
(Fig.~\ref{fig_cameradist}). Since the laser pulse height
was monitored continously during the measurement, a drift
of the laser signal can be excluded. A possible origin
are imperfections in the digital signal processing; 
the amplitude derived from the flash-ADC signals depends
slightly on the phase between the signal and the sampling 
clock. In principle, this effect is corrected, but residual
effects at the percent level are possible. Since the same effect
will occur for genuine Cherenkov signals, this small asymmetry
should not influence the quality of the calibration, which
refers to the average pulse height in both cases.
We nevertheless assign a systematic error
of 3\% corresponding to the difference between the mean and
the peak of the distribution.

Finally, there may be minor differences between the illumination
of the mirror with the light source at a distance of about
60~m, and with Cherenkov light from `infinity'. To ensure that
the mirror is uniformly illuminated, we rotated the
light source by $\pm 10^\circ$, equivalent to a displacement of 
the axis of the source from the telescope by three times the
mirror diameter; the amount of light detected at the telescope
was constant within 2-3\%. Differences in mirror obscuration
between the slightly divergent beam from the source, and ideal
parallel light are below 1\%. 

The resulting statistical and systematical errors are summarized
in Table~\ref{tab_errors}. To arrive at the final error given
in Table~\ref{tab_results}, the various systematic errors were
added in quadrature. The resulting total error is about 10\%.
Given the availability of photo sensors with 2\% calibration
errors, improved calibration procedures for the electronics,
and better temperature control,
we believe that with the further development of this technique
an overall 5\% error should be possible.
\begin{table} [htb]
\begin{center}
\begin{tabular}{|l|c|}
\hline
Error source & Error [\%] \\
\hline
Statistical errors & $< 0.4$ \\
Calibration of telescope photo sensor & 5.5 \\
Difference between calibration and operation cond. & 3.0 \\
Electronics gain, telescope photo sensor & 2.0 \\
Electronics gain, reference sensor, low gain (mode A) & 6.5 \\
Electronics gain, reference sensor, high gain (mode B) & 2.0 \\
Nonlinearity of camera PMs & 2.0 \\
Asymmetry of distribution of camera signals & 3.0 \\
Uniformity of mirror illumination and geometry & 1.5 \\
\hline
Total error & 10.2 \\
\hline
\end{tabular}
\vspace{0.5cm}
\caption{Sources of error in the determination of
$K_{IACT}$. The PM nonlinearity is only relevant for the
measurement at 430~nm, with its larger typical pulse heights.
For the total error, the different systematic
errors were added in quadrature.}
\label{tab_errors}
\end{center}
\end{table}

\section{Comparison with other calibration techniques}

It is of course interesting the see how the sensitivities
$K_{IACT}$ measured with the light source compare with 
other calibration techniques. 

We consider primarily the approach
where the individual factors contributing
to $K_{IACT}$ are estimated individually, as listed in
Table~\ref{tab_factors}.
The mirror area quoted includes the shadowing by the
camera masts etc.
The mirror reflectivity
is assumed to $(85 \pm 5) \%$ (compared to 89\% measured for new
mirrors~\cite{ct1}). 
Light collection by the camera PMs is governed
by a thin plexiglas  window (R\"ohm und Haas type 218, with improved
transparency at short wavelengths), and by
the funnels in front of the PMs.
Data-sheet values are used for the PMT quantum
efficiency, with a 15\% uncertainty on these values. 
Losses in photoelectron collection between the cathode and
the first dynode are neglected; since the funnels illuminate
only the central 15~mm of the PMTs, with their nominal cathode
diameter of 19~mm, such losses should be small.
The response of the readout to a 
single photoelectron is determined
based on the width of the laser test pulses used to flat-field
the camera, after corrections for the width of the single-photoelectron
response etc. 
Multiplying all these factors, one obtains the sensitivities 
$K_o$ given in Tables~\ref{tab_factors} and \ref{tab_results}, 
which are in good agreement with
the results from the optical calibration.
\begin{table} [htb]
\begin{center}
\begin{tabular}{|l|l|}
\hline
Mirror area              & $8.0 \pm 0.3$ m$^2$ \\
Mirror reflectivity      & $85 \pm 5$\% \\
Plexiglas camera cover   & $ 92 \pm 1$\% \\
Funnel light collectors  & $ 91 \pm 3$\% \\
Quantum efficiency       & $ 21 \pm 3$\% at 430 nm \\
                         & $ 15 \pm 2$\% at 470 nm \\
Conversion factor        & $0.95 \pm 0.14 $ ADC counts/photoel. \\
\hline
$K_o$                    & $(1.13 \pm 0.25) \cdot 10^4 $ ADC counts/(ph./cm$^2$)
                           at 430 nm \\
                         & $(0.81 \pm 0.18) \cdot 10^4 $ ADC counts/(ph./cm$^2$)
                           at 470 nm \\
\hline
\end{tabular}
\vspace{0.5cm}
\caption{Factors contributing to the estimate of $K_o$.}
\label{tab_factors}
\end{center}
\end{table}

A second approach uses the measured cosmic-ray trigger rate
of a telescope to derive its effective threshold and hence its
effective sensitivity. While final numbers are
still lacking, initial studies \cite{internal} indicate agreement
within the typical errors of about 15\%.

\section*{Acknowledgements}

The support of the German Ministry for Research 
and Technology BMBF is gratefully acknowledged. We thank the Instituto
de Astrofisica de Canarias for the use of the site and
for providing excellent working conditions. We have benefited
from discussions with other HEGRA members concerning telescope
calibration; E.~Lorenz, R.~Mirozyan and W.~Stamm
should be mentioned, as well as in particular F.~Aharonian, M.~Hemberger,
A.~Konopelko, M.~Panter and C.A. Wiedner.


\begin{thebibliography}{999}

\bibitem{overview} T.C.~Weekes, Space Science Rev. 75 (1996) 1;
M. F. Cawley and T.C.Weekes, Experimental Astronomy 6 (1996) 7.

\bibitem{cr_rejection} A.M.~Hillas, Space Science Rev. 75 (1996) 17.

\bibitem{stereo1} F.~Aharonian et al., Experimental Astronomy 2 (1993) 331.

\bibitem{stereo2} C.W.~Akerlof et al., Astrophys. J. 377 (1991) L97.

\bibitem{hegra_stereo} A.~Daum et al., preprint astro-ph/9704098 (1997),
Astroparticle Phys., in press; 
F.~Aharonian et al., preprint astro-ph 970619 (1997).

\bibitem{single_pe} R.~Mirzoyan, 
Proceedings of the Int. Workshop ``Towards a Major Atmospheric 
Cherenkov Detector IV'', Padua, (1995), M. Cresti (Ed.), p. 230.

\bibitem{npe} T.~Devlin et al., Nucl. Instr. meth. A268 (1988)24;
W.~Koska et al., FERMILAB-Pub-97/092 (1997).

\bibitem{npe1} A.G.~Wright, J. Phys. E 14 (1981) 851.

\bibitem{cr_cal} A.~Konopelko et al., Astroparticle Phys. 4 (1996) 199.

\bibitem{cr_flux} A.~Plyasheshnikov et al., submitted for publication.

\bibitem{cherenkov_rings} G. Vacanti et al., 
Astroparticle Phys. 2 (1994) 1.

\bibitem{starlight} O.~Karschnick, Diploma Theses, Kiel (1996).

\bibitem{ct1} R.~Mirzoyan et al., Nucl. Instr. Meth. A351 (1994) 513.

\bibitem{internal} F.~Aharonian, M.~Hemberger, A.~Konopelko,
internal note (1996), unpublished.

\end{thebibliography}
\end{document}